\documentclass[letterpaper, 10 pt, conference]{ieeeconf}  

\IEEEoverridecommandlockouts                              

\usepackage[latin1]{inputenc}																	%
\usepackage{type1cm}																			%
\usepackage{type1ec}																			%
\usepackage[T1]{fontenc}																		%
\usepackage{dsfont}																				%
\usepackage{booktabs}																			%
\usepackage{array}																				%
\usepackage{amsmath,amsfonts,amssymb,amstext}													%
\usepackage[amsmath,thmmarks,framed,hyperref]{ntheorem}									        %
\usepackage{tcolorbox} 																	        %
\usepackage{nicefrac}   								                                        %
\usepackage[]{units}																			%
\usepackage{enumerate}																			%
\usepackage{makeidx}																			%
\usepackage{fancyhdr}																			%
\usepackage{bm}																					%
\usepackage{lastpage}																			%
\usepackage{multicol}																			%
\usepackage{ifthen}																				%
\usepackage{ifpdf}												    							%
\usepackage{framed}																				%
\DeclareMathAlphabet{\mathpzc}{OT1}{pzc}{m}{it}													%
\usepackage{amsmath,amsfonts,amssymb,booktabs}												    %
\usepackage{psfrag} 																			%
\usepackage{float} 																				%
\usepackage{cite} 																				%
\usepackage{graphicx}                                                                           %
\usepackage{multirow}
\usepackage{nicematrix}
\usepackage{centernot}
\usepackage{nicefrac}                                                                           %
\usepackage{units}                                                                              %
\usepackage{graphicx} 																			%
\usepackage{textcomp} 																			%
\usepackage{color} 																				%
\usepackage{dsfont} 																			%
\usepackage{tikz}																				%
\usetikzlibrary{matrix}																			%
\usetikzlibrary{arrows,automata,matrix,fit,positioning,calc}         							%
\usepackage{cite}																				%
\usepackage{lipsum}                                                                             %
\usepackage[ruled,vlined]{algorithm2e}                                                          %
\makeatletter
\def\BState{\State\hskip-\ALG@thistlm}
\usepackage{xcolor,colortbl}                                                                    %
\usepackage{hhline}
\usepackage{hyperref}
\hypersetup{
    colorlinks=true,
    linkcolor=blue,
    filecolor=magenta,      
    urlcolor=cyan,
}
\usepackage{color, colortbl}
\definecolor{Gray}{rgb}{0.88,0.88,0.88}
\newcolumntype{g}{>{\columncolor{Gray}}c}

\newcommand{\x}{\boldsymbol{x}}
\newcommand{\bu}{\boldsymbol{u}}
\newcommand{\w}{\boldsymbol{w}}
\newcommand{\y}{\boldsymbol{y}}
\newcommand{\bv}{\boldsymbol{v}}
\newcommand{\e}{\boldsymbol{e}}

\newcounter{thmCounter}
\newtheorem{definition}[thmCounter]{\bfseries Definition}
\newtheorem{theorem}[thmCounter]{\bfseries Theorem}

\newtheorem{lemma}[thmCounter]{\bfseries Lemma}
\newtheorem{remark}[thmCounter]{\bfseries Remark}
\newtheorem{example}[thmCounter]{\bfseries Example}

\title{\LARGE \bf
Application of Monte Carlo Tree Search in Periodic Schedule Design for Networked Control Systems
}

\author{Burak Demirel and Arda Aytekin
\thanks{B. Demirel is with Ericsson Research, Torshamnsgatan 23, 164 40, Kista, Sweden
        {\tt\small burak.demirel@ericsson.com}.
		} 
\thanks{A. Aytekin is with Ericsson GAIA, Torshamnsgatan 23, 164 40, Kista, Sweden
        {\tt\small arda.aytekin@ericsson.com}.
		}
}

\begin{document}

\maketitle
\thispagestyle{empty}
\pagestyle{empty}

\begin{abstract}
We analyze the closed-loop control performance of a networked control system that consists of $N$ independent linear feedback control loops, sharing a communication network with $M$ channels ($M<N$). A centralized scheduler, employing a scheduling protocol that produces periodic communication sequences, dictates which feedback loops should utilize all these channels. Under the periodic scheduling protocol, we derive analytical expressions for quantifying the overall control performance of the networked control system in terms of a quadratic function. We also formulate the offline combinatorial optimization of communication sequences for a given collection of linear feedback control subsystems. Then, we apply Monte Carlo Tree Search to determine the period of these communication sequences that attain near-optimal control performance. Via numerical studies, we show the effectiveness of the proposed framework.
\end{abstract}

\begin{keywords}
    Networked Control Systems; Scheduling; Optimal Control; Monte Carlo Tree Search
\end{keywords}

\section{Introduction}\label{sec:intro}
As an ever-growing number of industrial devices become a part of the Internet-of-Things, the features, organizations, and operations of factories have radically changed. This radical change results in more flexible, continuous and flawless production than ever imagined before. Integrating industrial devices with computationally capable, embedded sensors and cutting-edge communication systems that enable ubiquitous and seamless connectivity allows us to monitor and operate industrial automation and control systems without any disruption at any time. However, the main challenge is to orchestrate a massive number of connected devices and machines in industrial control systems to attain an acceptable level of system performance.

As a result of Industry 4.0, the number of sensing and actuating elements connected to the 5G network in factories is exponentially growing~\cite{FVM+:19}. This abundance of devices share limited communication resources, such as time, frequency, space and energy. Therefore, there is an urgent need for efficiently deciding which sensors and actuators to address and what information to send at each time instant. In the literature, there exist a large variety of scheduling algorithms, which can be classified into two major groups: \emph{periodic}~\cite{HvK:02,LXF:03,ReS:04,HrZ:08,SCC:11,OBG:14} and \emph{aperiodic}~\cite{JoB:09,VZA+:12,GCH+:06,MGC+:11,HJT:12,ZCW+:18,BZC+:19,BZC+:21,DRQ+:18}. Periodic schedules are popular practical choices due to their low-implementation costs. The works~\cite{HvK:02} designed offline periodic scheduling policies to determine the stabilizing order of access to various sensors and/or actuators. The works~\cite{ReS:04,HrZ:08} studied linear-quadratic optimal control of multiple linear systems with limited communication. They formulated this optimal control problem as a combinatorial optimization that gives a solution to the optimal resource allocation problem. The work~\cite{LXF:03} determined an optimal periodic communication sequence and synthesized the associated optimal $H_{2}$ and $H_{\infty}$ controllers for networked control systems with limited communication resources. The works~\cite{SCC:11,OBG:14} focused on the sensor scheduling problem for estimation, wherein a collection of sensors share a common network to communicate their measurements. The work~\cite{OBG:14} proved that a Kalman-based scheduled filter produces periodic scheduling of the sensors. As an alternative, some works considered the design of aperiodic schedules from different views: stochastic scheduling~\cite{GCH+:06,MGC+:11}, event-triggered scheduling~\cite{HJT:12}, finite-horizon optimization~\cite{JoB:09,VZA+:12}, model predictive control approaches~\cite{ZCW+:18,BZC+:19,BZC+:21}, and reinforcement learning~\cite{DRQ+:18}.

\begin{figure}[!t] 
	\centering
    \includegraphics[scale=0.6]{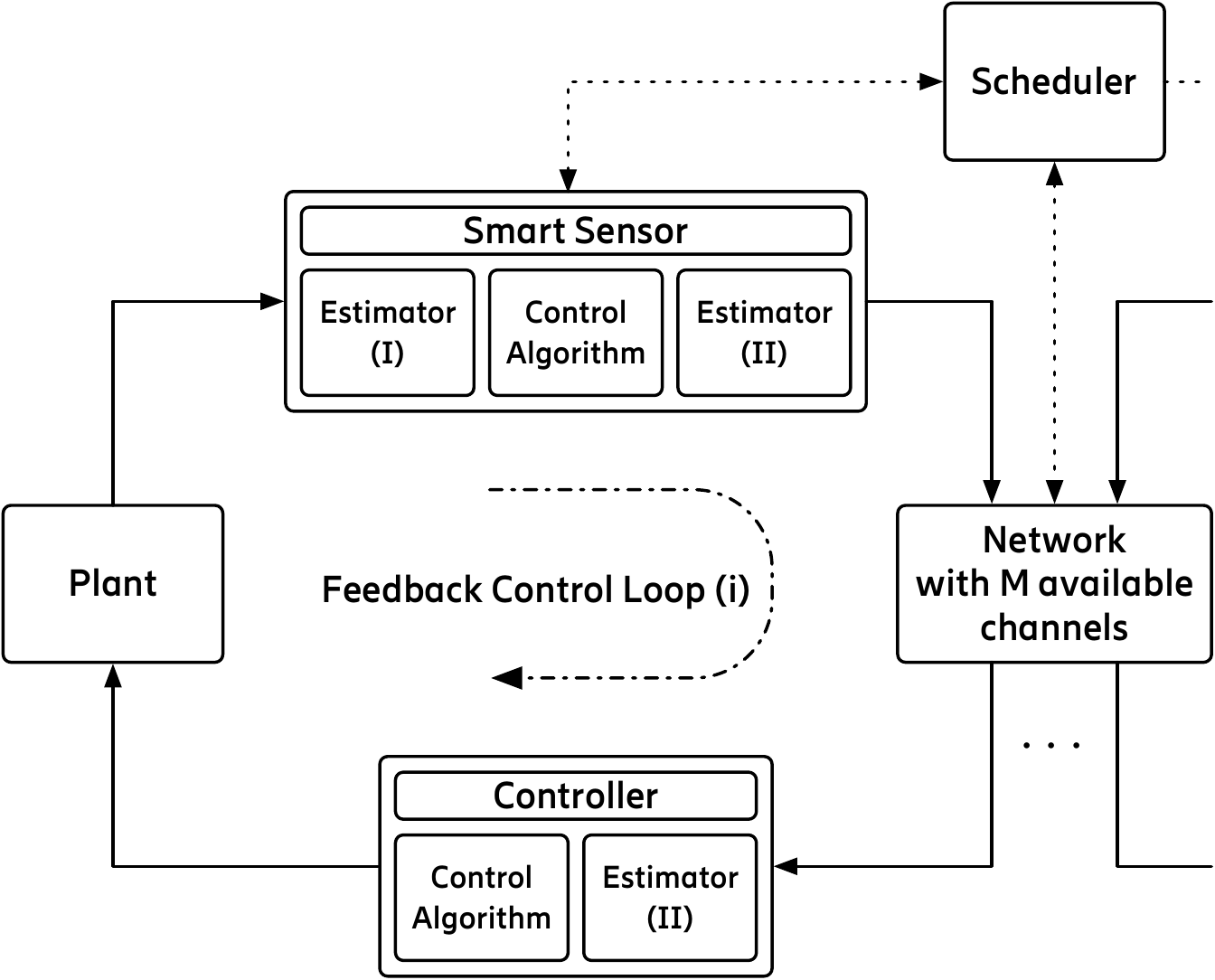}
	\caption{Networked control system that consists of $N$ feedback control loops closed over a shared medium with $M$ communication channels.}
	\label{fig:sys_arch}
\end{figure}

As mentioned earlier, finding the optimal communication sequence requires solving a combinatorial optimization problem. The computational effort, which needs to be spent to solve this optimization problem, explodes with an increasing number of sensors and actuators. Therefore, there is a dire need for a good heuristic to deal with large decision spaces. Monte Carlo tree search, introduced by Coulom~\cite{Cou:07}, is a popular technique for finding optimal decisions in planning problems by taking random samples in the decision space and constructing a search tree corresponding to the results~\cite{BPW+:12}. This technique is especially useful when dealing with large search spaces as it provides a way of intelligently exploring the domain by searching more promising parts of the search tree in more detail than the less promising ones.

\textbf{Contributions.} This paper considers a networked control system that consists of a collection of stochastic linear feedback control systems closed over a shared communication network with multiple channels; see Figure~\ref{fig:sys_arch}. Due to the lack of enough communication resources (i.e., the number of channels is strictly less than the number of feedback loops), in this paper, we use a centralized scheduler, which generates periodic communication sequences, to allocate available channels to feedback control loops. The main contributions of this paper are listed as follows.
\begin{itemize}
    \item[1)] We design the certainty-equivalent feedback controllers, which are optimal in our setup, since the scheduling decisions (that are made by the centralized scheduler) are independent of the control actions (that are computed by the controllers).
    \item[2)] We derive analytical expressions for quantifying the overall expected control loss in terms of a quadratic function.
    \item[3)] We use Monte Carlo Tree Search algorithm to find the periodic communication sequence that attains a \emph{near-optimal} performance. 
    \item[4)] We propose a flexible technique, which can efficiently scale up to an increasing number of plants and communication channels.
\end{itemize}

\textbf{Outline.} Section~\ref{sec:notations} provides notations used in the paper and the key definitions for the proof of the main theorem. Section~\ref{sec:networked_control_systems} introduces the main components of the networked control systems, together with necessary assumptions. Section~\ref{sec:comm_seq_design} presents Monte Carlo tree search for designing the near-optimal communication schedule. Numerical examples in Section~\ref{sec:numerical_example} highlight the power of our framework. Lastly, Section~\ref{sec:conclusion} provides concluding remarks while Appendix presents the proofs of Lemma~\ref{lem:eventually_periodic_sequence} and~\ref{lem:asymptotically_periodic_sequence}, and Theorem~\ref{thm:control_performance}.

\section{Notations and Preliminaries}\label{sec:notations}
We reserve $\mathbb{N}$ for the set of positive integers, $\mathbb{N}_{0}$ for $\mathbb{N}\cup\{0\}$, and $\mathbb{R}$ for the set of real numbers. We use $\mathbb{R}^{n}$ to denote the set of real vectors of dimension $n$. We write vectors in bold lower-case letters (e.g., $\bm{u}$ and $\bm{v}$) and matrices in capital letters (e.g., $A$ and $B$). The set of all real symmetric positive semi-definite matrices of dimension $n$ is represented by $\mathbb{S}_{\succeq 0}^{n}$. For a square matrix $A \in\mathbb{R}^{n\times n}$, $\mathrm{Tr}(A)$ denotes its trace, and $\lambda_{\max}(A)$ denotes its maximum eigenvalue in terms of magnitude. The notation $\{x_{k}\}_{k\in\mathcal{K}}$ stands for $\{x_{k} : k\in\mathcal{K}\}$, where $\mathcal{K}\subseteq\mathbb{N}_{0}$. We use $[N]$ to denote $\{1,\ldots,N\}$.

We review the essential definitions for building the results of this paper.

\begin{definition}[$\mu$-periodic sequence]
    A sequence $\{a_{t}^{}\}_{t\in\mathbb{N}_{0}}$ is $\mu$-periodic if $a_{t+\mu} = a_{t}$ for all $t\in\mathbb{N}_{0}$.
\end{definition}

\begin{definition}[Eventually $\mu$-periodic sequence]
    A sequence $\{a_{t}^{}\}_{t\in\mathbb{N}_{0}}$ is eventually $\mu$-periodic if there is an integer $t_{\circ}\geq 1$ such that $a_{t+\mu} = a_{t}$ for all $t_{\circ} \leq t\in\mathbb{N}_{0}$.
\end{definition}

Notice that discarding the first $t_{\circ}-1$ terms of an \emph{eventually periodic sequence} leads to a \emph{periodic sequence}.

\begin{definition}[Asymptotically $\mu$-periodic sequence] \label{def:asymptotically_periodic}
    A sequence $\{a_{t}^{}\}_{t\in\mathbb{N}_{0}}$ is asymptotically $\mu$-periodic if there exist two sequences $\{b_{t}^{}\}_{t\in\mathbb{N}_{0}}$ and $\{c_{t}^{}\}_{t\in\mathbb{N}_{0}}$ such that $\{b_{t}^{}\}_{t\in\mathbb{N}_{0}}$ is $\mu$-periodic, $\lim_{t\rightarrow\infty}c_{t}^{} = 0$, and $a_{t}^{} = b_{t}^{} + c_{t}^{}$ for all $t\in\mathbb{N}_{0}^{}$.
\end{definition}

\section{Networked Control Systems} \label{sec:networked_control_systems}
This section extensively reviews the networked control system architecture, illustrated in Figure~\ref{fig:sys_arch}, and its main components. In this section, we also introduce the essential assumptions under which we derive analytical expressions, provided in Section~\ref{subsec:controllers}, for the closed-loop control performance.

\subsection{Control system architecture}\label{subsec:architecture}
As illustrated in Figure~\ref{fig:sys_arch}, we here consider a networked control system that consists of $N$ independent feedback loops closed over a shared communication network that comprises $M$ communication channels. Since the number of communication channels is strictly less than the number of subsystems, only a subset of feedback loops can be closed at each sampling interval. Therefore, a centralized scheduler orchestrates communication among entities (i.e., sensors and controllers) of these feedback control loops.

Each feedback control loop consists of a smart sensor, a controller, and an actuator. As depicted in Figure~\ref{fig:sys_arch}, each controller is collocated with an actuator but not with a sensor. Each sensor periodically takes noisy measurements of the subsystem's output at a fixed sampling rate. Then, each sensor computes the state estimates based on its measurements and transmits them to an associated remote controller whenever the scheduler allocates an available channel to this sensor. Each remote controller computes the control commands based on either its estimates or the sensor's estimates (depending on the scheduler's decision) and sends the commands immediately to the actuator. Each actuator acts whenever it receives control commands. All data transmissions that take place in the networked control system are immediate and lossless.

\subsection{Plants}\label{subsec:plants}
We consider a group of linear time-invariant discrete-time stochastic systems, i.e.,
\begin{equation}
    \begin{split}
	    \x_{t+1}^{(i)} =&\; A_{}^{(i)}\x_{t}^{(i)} + B_{}^{(i)}\bu_{t}^{(i)} + \w_{t}^{(i)} \;, \\
	    \y_{t}^{(i)} =&\; C_{}^{(i)}\x_{t}^{(i)} + \bv_{t}^{(i)} \;, 
    \end{split} \label{eqn:LTI_system}
\end{equation}
where $\x_{t}^{(i)}\in\mathbb{R}_{}^{n_i}$, $\bu_{t}^{(i)}\in\mathbb{R}_{}^{m_i}$ and $\y_{t}^{(i)}\in\mathbb{R}_{}^{p_i}$ for all $i\in [N]$ denote the subsystem $i$'s state, control input and output, respectively, at any $t\in\mathbb{N}_{0}^{}$. We assume that the noise sources, $\w_{t}^{(i)}\in\mathbb{R}_{}^{n_i}$ and $\bv_{t}^{(i)}\in\mathbb{R}_{}^{p_i}$, are uncorrelated zero-mean i.i.d. Gaussian random vectors with covariance matrices $W_{}^{(i)}$ and $V_{}^{(i)}$, respectively. The initial state of the subsystem $i$, $\x_{0}^{(i)}$, is assumed to be a Gaussian random vector with mean $\bar{\x}_{0}^{(i)}$ and covariance matrix $X_{0}^{(i)}$. All noise sources, $\w_{t}^{(i)}$ and $\bv_{t}^{(i)}$, are independent of the initial conditions $\x_{0}^{(i)}$.

\subsection{Smart sensors and pre-processing units}
In our setup, smart sensors, which have both computing and communication capabilities, play a central role. Each sensor not only samples the subsystem's output periodically but also computes the state estimates by using a standard Kalman filter (i.e., Estimator I in Figure~\ref{fig:sys_arch}). The sensor sends its estimates to the controller instead of the raw measurements if the scheduler allocates an available channel to this sensor for data dissemination. In case the controller does not receive any updated information of the state estimate, it uses its own estimator (i.e., Estimator II in Figure~\ref{fig:sys_arch}) to compute the state estimate based on an open-loop system model. The sensor runs a copy of the estimator on the controller side (i.e., Estimator II in Figure~\ref{fig:sys_arch}) together with an identical control algorithm implemented in the controller to compute control commands applied by the actuator to the plant. We now review the estimators implemented in the sensors and the controllers. 

\noindent\textbf{Estimator (I).} The smart sensor utilizes a standard Kalman filter to calculate the state estimate $\hat{\x}_{t \mid t}^{s(i)}$ and covariance $P_{t\mid t}^{s(i)}$ recursively as
\begin{align*}
	\hat{\x}_{t \mid t-1}^{s(i)} &= A_{}^{(i)}\hat{\x}_{t-1 \mid t-1}^{s(i)} + B_{}^{(i)}\bu_{t-1}^{(i)} \\
	P_{t \mid t-1}^{s(i)} &= A_{}^{(i)}P_{t-1 \mid t-1}^{s(i)}A_{}^{(i)\top} + W_{}^{(i)} \\
	K_{t}^{} &=  P_{t\mid t-1}^{s(i)}C_{}^{(i)\top}\left( C_{}^{(i)}P_{t\mid t-1}^{s(i)}C_{}^{(i)\top} + V_{}^{(i)} \right)_{}^{-1} \\
	\hat{\x}_{t \mid t}^{s(i)} &= \hat{\x}_{t \mid t-1}^{s(i)} + K_{t}^{(i)}\left( \y_{t}^{(i)} - C\hat{\x}_{t \mid t-1}^{s(i)} \right) \\
	P_{t\mid t}^{s(i)} &= \left( \mathbf{I}_{n_{i}} - K_{t}^{(i)}C_{}^{(i)} \right) P_{t\mid t-1}^{s(i)} \;,
\end{align*}
starting from $\hat{\x}_{0 \mid -1}^{s(i)} = \bar{\x}_{0}^{(i)}$ and $P_{0 \mid -1}^{s(i)} = X_{0}^{(i)}$.

\noindent\textbf{Estimator (II).} The feedback controller runs an estimator to compute the state estimate $\hat{\x}_{t \mid t}^{c(i)}$ as
\begin{align}
	\hat{\x}_{t \mid t-1}^{c(i)} & =  A_{}^{(i)} \hat{\x}_{t-1 \mid t-1}^{c(i)} + B_{}^{(i)} \bu_{t-1}^{(i)} \;, \label{eqn:prediction_step_controller} \\
	\hat{\x}_{t \mid t}^{c(i)} & =
	\begin{cases}
		\hat{\x}_{t \mid t}^{s(i)} & \text{if the MMSE estimate received}\;, \\
		\hat{\x}_{t \mid t-1}^{c(i)} & \text{otherwise} \;,
	\end{cases} \label{eqn:update_step_controller} 
\end{align}
with $\hat{\x}_{0 \mid -1}^{c(i)} = \bar{\x}_{0}^{(i)}$.

We have $\hat{\x}_{t \mid t}^{c(i)} =  \hat{\x}_{t \mid t}^{s(i)}$ when the sensor and controller of the $i^{\text{th}}$ feedback loop have communicated. Otherwise, $\hat{\x}_{t \mid t}^{c(i)}=\hat{\x}_{t \mid t-1}^{c(i)}$, the state estimate obtained from Estimator (II).

\subsection{Scheduler}
We employ a centralized scheduler that orchestrates communication over a shared medium, which can only accommodate a maximum number of feedback control loops (i.e., $M$ out of $N$ feedback loops) at a time. The scheduler, therefore, generates a $T_{0}$-\emph{periodic communication sequence} defined by
\begin{align*}
    \left\{ \sigma_{t}^{(i)}\in\{0,1\} : \sum_{i=1}^{N} \sigma_{t}^{(i)}=M, ~\sigma_{t+T_{0}}^{(i)}=\sigma_{t}^{(i)},~\forall t\in\mathbb{N}_{0}^{} \right\}
\end{align*}
for all $i\in [N]$ to decide which $M$ of the $N$ feedback loops are allocated $M$ available channels at any $t\in\mathbb{N}_{0}^{}$. Notice that $\sigma_{t}^{(i)}$ are binary decisions that indicate whether the $i^{\mathrm{th}}$ feedback loop is closed ($\sigma_{t}^{(i)}=1$), or not ($\sigma_{t}^{(i)}=0$).

 The scheduler's decisions determine the \emph{elapsed time} since the last transmission of sensor packets for all feedback loops. We introduce an integer-valued variable, $\tau_{t}^{(i)}$, to describe the elapsed time for the $i^{\mathrm{th}}$ subsystem. The evolution of this variable is defined by
\begin{align}
	\tau_{t}^{(i)}  =
	\begin{cases}
		0 & \text{if} ~ \sigma_{t}^{(i)} = 1 \;, \\
		1 + \tau_{t-1}^{(i)} & \text{otherwise} \;,
	\end{cases} 
	\label{eqn:timer}
\end{align}
where $\tau_{t}^{(i)} = 0$ for all $t<0$. Notice that, for a given $i\in [N]$, if there exists at least one $m\in\{0, 1, \cdots, T_{0}-1\}$ such that $\sigma_{m+kT_{0}}^{(i)}=1, ~\forall k\in\mathbb{N}_{0}$, then the number of time steps between two consecutive transmissions for the $i^{\mathrm{th}}$ subsystem is bounded. Otherwise, it becomes unbounded. 

\begin{lemma}\label{lem:eventually_periodic_sequence}
    Suppose that $\{ \sigma_{t}^{(i)} \}_{t\in\mathbb{N}_{0}}$ is a $T_{0}$-periodic binary sequence and there exists at least one $m\in\{0, 1, \cdots, T_{0}-1\}$ such that $\sigma_{m+kT_{0}}^{(i)}=1, ~\forall k\in\mathbb{N}_{0}$. If $\sigma_{k T_{0}}^{(i)}=\sigma_{(k+1)T_{0}-1}^{(i)}=0, ~\forall k\in\mathbb{N}_{0}$, then $\{ \tau_{t}^{(i)} \}_{t\in\mathbb{N}_{0}}$ is an eventually $T_{0}$-periodic sequence of integers that are strictly less than $T_{0}$. Otherwise, $\{ \tau_{t}^{(i)} \}_{t\in\mathbb{N}_{0}}$ is a $T_{0}$-periodic sequence of integers that are strictly less than $T_{0}$.
\end{lemma}

\begin{remark}\label{rem:periodic_sequence}
    If $\{ \sigma_{t}^{(i)} \}_{t\in\mathbb{N}_{0}}$ is a $T_{0}$-periodic binary sequence and there exists at least one $m\in\{0, 1, \cdots, T_{0}-1\}$ such that $\sigma_{m+kT_{0}}^{(i)}=1, ~\forall k\in\mathbb{N}_{0}$, then $\{ \tau_{t}^{(i)} \}_{T_{0}\leq t\in\mathbb{N}_{0}}$ is a $T_{0}$-periodic sequence of integers that are strictly less than $T_{0}$. 
\end{remark}

The following example provides a better understanding of how the parameters $\sigma_{t}^{(i)}$ and $\tau_{t}^{(i)}$ evolve over time.

\begin{example}
    The networked control system, illustrated in Figure~\ref{fig:sys_arch}, consists of three independent feedback loops closed over a network with two communication channels. At each sampling instant, the centralized scheduler allocates these available channels to two out of three feedback control loops. The scheduler, therefore, generates two periodic channel allocation sequences with a period of five, i.e., $\{1,2,3,1,2\}$ and $\{2,3,1,3,1\}$. To achieve these channel allocation sequences, we form periodic decision sequences (shown in Table~\ref{tab:example}) with a period of five; for Sensor 1, 2, and 3, as $\{1,0,1,1,1\}$, $\{1,1,0,0,1\}$, and $\{0,1,1,1,0\}$, respectively. Thus, the sequences of the elapsed time since the last transmission of the sensor packet in feedback loops 1 and 2 are periodic, i.e., $\{ 0,1,0,0,0 \}$ and $\{ 0,0,1,2,0 \}$. However, as seen in the last row of Table~\ref{tab:example}, the sequence of the elapsed time since the last transmission of the sensor packet in feedback loop 3 is eventually periodic because $\{1,0,0,0,1,2,0,0,0,1,\ldots\}$ is periodic after the fifth time instant (i.e., first period).
\end{example}

\begin{table}[ht!]
\caption{Example: The evolution of $\sigma_{t}^{(i)}$ and $\tau_{t}^{(i)}$}
\centering
\begin{tabular}{c g g g g g c c c c c g}
\toprule
$t$ & 0 & 1 & 2 & 3 & 4 & 5 & 6 & 7 & 8 & 9 & \ldots \\
\midrule
$\sigma_{t}^{(1)}$  & 1 & 0 & 1 & 1 & 1 & 1 & 0 & 1 & 1 & 1 & \ldots \\ 
$\sigma_{t}^{(2)}$  & 1 & 1 & 0 & 0 & 1 & 1 & 1 & 0 & 0 & 1 & \ldots \\ 
$\sigma_{t}^{(3)}$  & 0 & 1 & 1 & 1 & 0 & 0 & 1 & 1 & 1 & 0 & \ldots \\ 
\midrule
$\tau_{t}^{(1)}$    & 0 & 1 & 0 & 0 & 0 & 0 & 1 & 0 & 0 & 0 & \ldots \\ 
$\tau_{t}^{(2)}$    & 0 & 0 & 1 & 2 & 0 & 0 & 0 & 1 & 2 & 0 & \ldots \\ 
$\tau_{t}^{(3)}$    & 1 & 0 & 0 & 0 & 1 & 2 & 0 & 0 & 0 & 1 & \ldots \\ 
\bottomrule
\end{tabular}
\label{tab:example}
\end{table}

As can be perceived by examining~\eqref{eqn:update_step_controller}, the decisions of the scheduler have an influence on the estimation error on the controller side. We define the estimation error on the controller node by 
\begin{align}
    \tilde{\x}_{t\mid t}^{c(i)} = \x_{t}^{(i)} - \hat{\x}_{t\mid t}^{c(i)} \;,
\end{align}
which evolves as
\begin{align}
    \tilde{\x}_{t\mid t}^{c(i)} = 
	\begin{cases}
		\tilde{\x}_{t\mid t}^{s(i)} & \text{if} ~ \sigma_{t}^{(i)} = 1 \;, \\
		A_{}^{(i)}\tilde{\x}_{t-1\mid t-1}^{c(i)} + \w_{t-1}^{(i)} & \text{otherwise} \;.
	\end{cases} 
\end{align}
Similarly, the state estimation error between the sensor and the controller can be defined by
\begin{align}
    \e_{t\mid t}^{(i)} = \hat{\x}_{t \mid t}^{s(i)} - \hat{\x}_{t \mid t}^{c(i)}\;, \label{eqn:error_sensor_controller}
\end{align}
which evolves as
\begin{align}
    \e_{t\mid t}^{(i)} = 
	\begin{cases}
		0 & \text{if} ~ \tau_{t}^{(i)} = 0 \;, \\
		A_{}^{(i)}\e_{t-1\mid t-1}^{(i)} + \boldsymbol{\eta}_{t-1}^{(i)} & \text{otherwise} \;,
	\end{cases} \label{eqn:error_covariance_evolution}
\end{align}
where $\boldsymbol{\eta}_{t}^{(i)}\in\mathbb{R}^{n_{i}}$ is a zero-mean i.i.d. Gaussian random vector with covariance matrix $\Pi_{t}^{(i)} \triangleq K_{t}^{(i)}C_{}^{(i)}P_{t\mid t}^{s(i)}$; see~\cite{DLG+:19}.

The covariance of the random variable $\e_{t\mid t}^{(i)}$ evolves as
\begin{align}\label{eqn:covariance}
	\Sigma_{t}^{(i)} =  
	\begin{cases}
		\mathbf{0}_{n_{i}} & \text{if} ~ \tau_{t}^{(i)} = 0 \;, \\
		A_{}^{(i)}\Sigma_{t-1}^{(i)}A_{}^{(i)\top} + \Pi_{t-1}^{(i)} & \text{otherwise} \;,
	\end{cases} 
\end{align}
where $\Sigma_{t}^{(i)} = \mathbf{0}_{n_{i}}$ for all $t<0$.

\begin{lemma}\label{lem:asymptotically_periodic_sequence}
    Suppose that $\{ \sigma_{t}^{(i)} \}_{t\in\mathbb{N}_{0}}$ is a $T_{0}$-periodic binary sequence. If there exists at least one $m\in\{0, 1, \cdots, T_{0}-1\}$ such that $\sigma_{m+kT_{0}}^{(i)}=1, ~\forall k\in\mathbb{N}_{0}$, then $\{ \Sigma_{t}^{(i)} \}_{t\in\mathbb{N}_{0}}$ is asymptotically $T_{0}$-periodic.
\end{lemma}

We want to stress that Lemma~\ref{lem:asymptotically_periodic_sequence} has a pivotal role in deriving analytical expressions for the minimum expected infinite-horizon control loss. 

\subsection{Controllers and cost functions}\label{subsec:controllers}
Under periodic scheduling decisions made by the centralized scheduler, we want to compute a set of control commands to minimize the overall control loss of the form
\begin{align}\label{eq_control_loss}
	J_{T} = \sum_{i = 1}^{N} J_{T}^{(i)},
\end{align}
where $J_{T}^{(i)}$ is the control loss of the $i^{\mathrm{th}}$ subsystem and is given by
\begin{multline}
	J_{T}^{(i)} = \mathop{\mathbf{E}}\bigg[ \x_{T}^{(i)\top}Q_{f}^{(i)}\x_{T}^{(i)} \\ + \sum_{t=0}^{T-1}\Big( \x_{t}^{(i)\top}Q_{}^{(i)}\x_{t}^{(i)} + \bu_{t}^{(i)\top}R_{}^{(i)}\bu_{t}^{(i)} \Big) \bigg] \;,
	\label{eqn:control_loss_function}
\end{multline}
where $Q_{}^{(i)}\in\mathbb{S}_{\succeq 0}^{n}$, $Q_{f}^{(i)}\in\mathbb{S}_{\succeq 0}^{n}$ and $R_{}^{(i)}\in\mathbb{S}_{\succ 0}^{n}$ are the state, terminal and control weight matrices, respectively, subject to the dynamics~\eqref{eqn:LTI_system}.

At any time $t\in\mathbb{N}_{0}$, the \emph{certainty equivalent controller}, which is optimal under exogenous schedules~\cite{MoH:13, DLG+:19}, in the $i^{\mathrm{th}}$ feedback loop computes control actions, based upon
\begin{align} \label{eq_control_action}
	\bu_{t}^{(i)} = - L_{t}^{(i)} \hat{\x}_{t \mid t}^{c(i)},
\end{align}
where $\hat{\x}_{t \mid t}^{c(i)}$ is the state estimate used by the controller,
\begin{align}
	L_{t}^{(i)} = (B_{}^{(i)\top}S_{t+1}^{(i)}B_{}^{(i)} + R_{}^{(i)})_{}^{-1} B_{}^{(i)\top}S_{t+1}^{(i)}A_{}^{(i)},  \label{eqn:optimal_control_gain}
\end{align}
and $S_{t}^{(i)}$ is recursively computed as
\begin{multline}
	S_{t}^{(i)} = A_{}^{(i)\top} S_{t+1}^{(i)} A_{}^{(i)} + Q_{}^{(i)} - A_{}^{(i)\top}S_{t+1}^{(i)}B_{}^{(i)} \\ \times(B_{}^{(i)\top}S_{t+1}^{(i)}B_{}^{(i)} + R_{}^{(i)})_{}^{-1} B_{}^{(i)\top}S_{t+1}^{(i)}A_{}^{(i)} , \label{eqn:riccati_equation}
\end{multline}
with initial condition $S_{T}^{(i)} = Q_{f}^{(i)}$. The minimum value of the control loss of the $i^{\mathrm{th}}$ subsystem is
\begin{multline}
		J_{T}^{(i)} =\; \bar{\x}_{0}^{(i)\top}S_{0}^{(i)}\bar{\x}_{0}^{(i)} + \textnormal{Tr}\big( S_{0}^{(i)}X_{0}^{(i)} \big) + \sum_{t=0}^{T-1}\textnormal{Tr}\big( S_{t+1}^{(i)}W_{}^{(i)} \big) \\
		+ \sum_{t=0}^{T-1} \textnormal{Tr}\big( P_{t \mid t}^{s(i)} \Gamma_{t}^{(i)} \big) + \sum_{t=0}^{T-1} \mathbf{E}\Big[ \e_{t \mid t}^{(i)\top} \Gamma_{t}^{(i)} \e_{t \mid t}^{(i)} \Big] \;, \label{eqn:inf_expected_cost} 
\end{multline}
where $\Gamma_{t}^{(i)} \triangleq L_{t}^{(i)\top}(B_{}^{(i)\top}S_{t+1}^{(i)}B_{}^{(i)} + R_{}^{(i)})L_{t}^{(i)}$ and $\e_{t \mid t}^{(i)} \triangleq \hat{\x}_{t \mid t}^{s(i)} - \hat{\x}_{t \mid t}^{c(i)}$. 

\begin{theorem}\label{thm:control_performance}
    Let $\{\sigma_{t}^{(i)}\}_{t\in\mathbb{N}_{0}}$ be a $T_{0}$-periodic binary sequence. Suppose that $(A_{}^{(i)},B_{}^{(i)})$ and $(A_{}^{(i)},W_{}^{(i)\nicefrac{1}{2}})$ are controllable, and $(A_{}^{(i)},C_{}^{(i)})$ and $(A_{}^{(i)},Q_{}^{(i)\nicefrac{1}{2}})$ are observable for all $i\in [N]$. Then, for any $i\in [N]$ when $T\rightarrow\infty$, the following statements are true:
    \begin{enumerate}[(i)]
        \item The matrices $S_{\infty}^{(i)}$ and $P_{\infty}^{s(i)}$ are the positive definite solutions of the following algebraic Riccati equations:
    \begin{align*}
	    &S_{\infty}^{(i)} = A_{}^{(i)\top} S_{\infty}^{(i)} A_{}^{(i)} + Q_{}^{(i)} - A_{}^{(i)\top}S_{\infty}^{(i)}B_{}^{(i)} \nonumber\\ &\times\left(B_{}^{(i)\top}S_{\infty}^{(i)}B_{}^{(i)} + R_{}^{(i)}\right)_{}^{-1} B_{}^{(i)\top}S_{\infty}^{(i)}A_{}^{(i)}, \\
	    &P_{\infty}^{s(i)} = A_{}^{(i)} P_{\infty}^{s(i)} A_{}^{(i)\top} + W_{}^{(i)} - A_{}^{(i)}P_{\infty}^{s(i)}C_{}^{(i)\top} \nonumber\\ &\times\left(C_{}^{(i)}P_{\infty}^{s(i)}C_{}^{(i)\top} + V_{}^{(i)}\right)_{}^{-1} C_{}^{(i)}P_{\infty}^{s(i)}A_{}^{(i)\top}.
    \end{align*}
        \item The optimal control gain becomes constant, i.e.,
        \begin{align*}
            L_{\infty}^{(i)} = \left(B_{}^{(i)\top}S_{\infty}^{(i)}B_{}^{(i)} + R_{}^{(i)}\right)_{}^{-1}B_{}^{(i)\top}S_{\infty}^{(i)}A_{}^{(i)}.
        \end{align*}
        \item The optimal estimation gain becomes constant, i.e.,
        \begin{align*}
            K_{\infty}^{(i)} = P_{\infty}^{s(i)}C_{}^{(i)\top}\left( C_{}^{(i)}P_{\infty}^{s(i)}C_{}^{(i)\top} + V_{}^{(i)} \right)_{}^{-1}.
        \end{align*}
        \item If there exists at least one $m\in\{0, 1, \cdots, T_{0}-1\}$ such that $\sigma_{m+kT_{0}}^{(i)}=1, ~\forall k\in\mathbb{N}_{0}$, then the minimum expected control loss converges to
        \begin{multline}
            J_{\textrm{ave}}^{(i)} \triangleq  \lim_{T\rightarrow\infty}\frac{1}{T}J_{T}^{(i)} = \mathrm{Tr}\big( S_{\infty}^{(i)}W_{}^{(i)} \big) \\ + \mathrm{Tr}\big( F_{\infty}^{s(i)}\Gamma_{\infty}^{(i)} \big)
    + \frac{1}{T_{0}^{}}\sum_{t=T_{0}}^{2T_{0}-1}\mathrm{Tr}\big( \Gamma_{\infty}^{(i)}\bar{\Sigma}_{t}^{(i)} \big), \label{eqn:infinite_horizon_cost}
        \end{multline}
        with, starting from $\bar{\Sigma}_{t}^{(i)}=\mathbf{0}_{n_{i}}$ for all $t<\infty$,
        \begin{align*}
        \bar{\Sigma}_{t}^{(i)} =  
	        \begin{cases}
		        \mathbf{0}_{n_{i}} & \text{if} ~ \sigma_{t}^{(i)} = 1 \;, \\
		A_{}^{(i)}\bar{\Sigma}_{t-1}^{(i)}A_{}^{(i)\top} + \Pi_{\infty}^{(i)} & \text{otherwise} \;,
	        \end{cases} 
        \end{align*}
        where $F_{\infty}^{s(i)}\triangleq \left( \mathbf{I}_{n_{i}}^{} - K_{\infty}^{(i)}C_{}^{(i)} \right)P_{\infty}^{s(i)}$ and $\Gamma_{\infty}^{(i)}\triangleq L_{\infty}^{(i)\top}\left(B_{}^{(i)\top}S_{\infty}^{(i)}B_{}^{(i)} + R_{}^{(i)} \right)L_{\infty}^{(i)}$.

        Otherwise, 
        \begin{itemize}
            \item If $\lambda(A^{(i)}) < 1$, then the minimum expected control loss converges to
            \begin{multline}
                J_{\textrm{ave}}^{(i)} \triangleq  \lim_{T\rightarrow\infty}\frac{1}{T}J_{T}^{(i)} = \mathrm{Tr}\big( S_{\infty}^{(i)}W_{}^{(i)} \big) \\ + \mathrm{Tr}\big( F_{\infty}^{s(i)}\Gamma_{\infty}^{(i)} \big)
                + \mathrm{Tr}\big( Z_{\infty}^{(i)}\Gamma_{\infty}^{(i)} \big) \;,
            \end{multline}
            where $Z_{\infty}^{(i)}$ are the positive semi-definite solutions of the following equations:
            \begin{align*}
                A^{(i)}Z_{\infty}^{(i)}A^{(i)\top} -  Z_{\infty}^{(i)} + \Pi_{\infty}^{(i)} = 0 \;.
            \end{align*}
            \item If $\lambda(A^{(i)}) \geq 1$, then the minimum expected control loss diverges, i.e., $J_{\textrm{ave}}^{(i)} = +\infty$.
        \end{itemize}
    \end{enumerate}
\end{theorem}

\begin{figure}[t]
\centering
\includegraphics[scale=0.52]{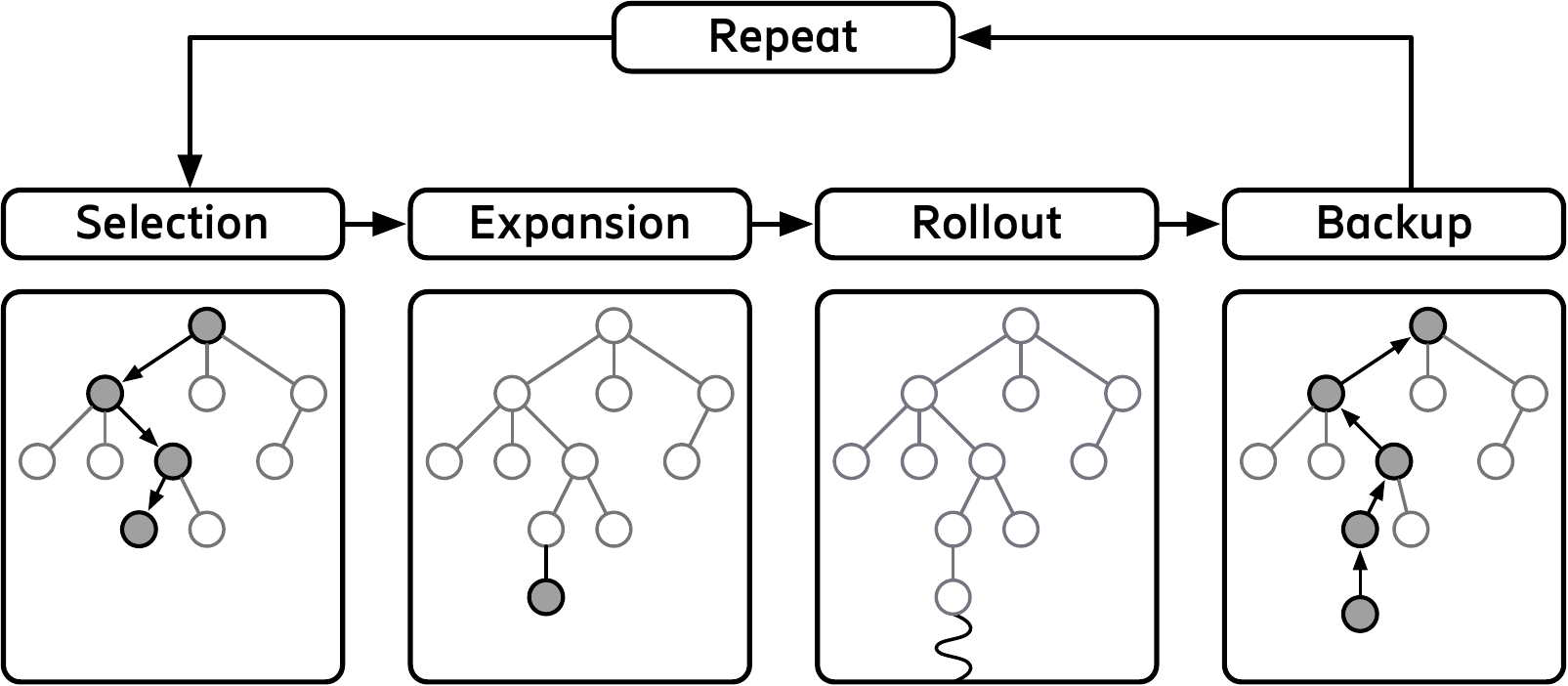}
\caption{Monte Carlo Tree Search. Each iteration consists of the four distinct steps: Selection, Expansion, Roll-out, and Backup.}\label{fig:MCTS}
\end{figure}

\section{Communication Sequence Design via Monte Carlo Tree Search} \label{sec:comm_seq_design}
The MCTS builds a tree starting from a root node (i.e., \emph{an empty set} since none of the channels are allocated in the beginning) in an incremental fashion. Each node of the tree represents the channels' allocation to a subset of feedback loops and records statistics concerning its children (i.e., visit counts and accumulated values). The algorithm runs for a certain number of iterations from the root node (i.e., an empty set) and, in each iteration, it repeatedly executes four distinct steps (see Figure~\ref{fig:MCTS}) listed below.
\begin{itemize}
    \item {\bf Selection.} The current grown tree is traversed starting from the root node until reaching a node which is not expanded. During this phase, the child is selected via \emph{Upper Confidence Bounds applied to Trees}, given by
    \begin{align}
        \arg\max_{a\in\mathcal{A}(s)} \frac{\mathbf{W}(s,a)}{\mathbf{N}(s,a)} + c_{uct}\sqrt{\frac{\log \mathbf{N}(s)}{\mathbf{N}(s,a)}}\;,
    \end{align}
    where $\mathbf{N}(s,a)$ is the number of times action $a$ has been selected at state $s$, $\mathbf{W}(s,a)$ is the cumulative sum of returns when taking action $a$ in state $s$, $\mathbf{N}(s)\triangleq\sum_{a\in\mathcal{A}(s)}\mathbf{N}(s,a)$ is the number of times node $s$ has been visited, and $c_{uct}>0$ is the constant striking a balance between exploration and exploitation.
    \item {\bf Expansion.} The tree is expanded by adding a new node as a child to the leaf node.
    \item {\bf Roll-out.} The value of the new node is computed by repeatedly choosing random actions from that node until reaching a terminal node, and taking the outcome.
    \item {\bf Backup.} The outcome of the roll-out phase is propagated up to all nodes encountered in the selection phase by updating their respective statistics as
    \begin{align*}
        \mathbf{W}(s,a) &\leftarrow \mathbf{W}(s,a) + (1-\nicefrac{J_{\textrm{ave}}}{J_{\max}}) \;, \\
        \mathbf{N}(s,a) &\leftarrow \mathbf{N}(s,a) + 1 \;.
    \end{align*}
\end{itemize}

After running out of the computational budget, the MCTS algorithm returns a sequence of nodes from the root to a leaf, which provides the best value encountered so far. The sequence of nodes corresponds to the sequence of communication allocations.

\begin{table*}[ht!]
\caption{Optimal periodic schedules for three feedback loops while sharing a single channel}
\centering
\begin{tabular}{| c | c | >{\centering}p{0.4cm} | >{\centering}p{0.4cm} | >{\centering}p{0.4cm} | >{\centering}p{0.4cm} | >{\centering}p{0.4cm} | >{\centering}p{0.4cm} | >{\centering}p{0.4cm} | >{\centering}p{0.4cm} | >{\centering}p{0.4cm} | >{\centering}p{0.4cm} | >{\centering}p{0.4cm} | >{\centering}p{0.4cm} | c |}
\hline
\multirow{3}{*}{Period $T_{0}$} & \multirow{3}{*}{Plant index $i$} & \multicolumn{12}{l|}{Periodic sequence $\big\{\sigma_{m + kT_{0}}^{(i)}\big\}_{k\in\mathbb{N}_{0}}, ~ 0\leq m <T_{0}$} & \multirow{3}{*}{Total loss $J_{\textrm{ave}}$} \\ \cline{3-14} 
& & \multicolumn{12}{c|}{$m$} & \\ \cline{3-14}
& & 0 & 1 & 2 & 3 & 4 & 5 & 6 & 7 & 8 & 9 & 10 & 11 & \\ \hline
\multirow{3}{*}{3}  & 1 & 0 & 0 & 1 & \multicolumn{9}{l|}{}                                     & \multirow{3}{*}{$576.2013$} \\ \cline{2-5} 
                    & 2 & 1 & 0 & 0 & \multicolumn{9}{l|}{}                                     &                           \\ \cline{2-5} 
                    & 3 & 0 & 1 & 0 & \multicolumn{9}{l|}{}                                     &                           \\ \cline{1-6}\cline{15-15}
\multirow{3}{*}{4}  & 1 & 0 & 1 & 0 & 1 & \multicolumn{8}{l|}{}                                 & \multirow{3}{*}{$385.9708$} \\ \cline{2-6} 
                    & 2 & 1 & 0 & 0 & 0 & \multicolumn{8}{l|}{}                                 &                           \\ \cline{2-6} 
                    & 3 & 0 & 0 & 1 & 0 & \multicolumn{8}{l|}{}                                 &                           \\ \cline{1-7}\cline{15-15}
\multirow{3}{*}{5}  & 1 & 0 & 1 & 0 & 1 & 1 & \multicolumn{7}{l|}{}                             & \multirow{3}{*}{$399.8308$} \\ \cline{2-7} 
                    & 2 & 1 & 0 & 0 & 0 & 0 & \multicolumn{7}{l|}{}                             &                           \\ \cline{2-7} 
                    & 3 & 0 & 0 & 1 & 0 & 0 & \multicolumn{7}{l|}{}                             &                           \\ \cline{1-8}\cline{15-15}
\multirow{3}{*}{6}  & 1 & 0 & 1 & 0 & 1 & 0 & 1 & \multicolumn{6}{l|}{}                         & \multirow{3}{*}{$385.3658$} \\ \cline{2-8} 
                    & 2 & 1 & 0 & 1 & 0 & 0 & 0 & \multicolumn{6}{l|}{}                         &                           \\ \cline{2-8} 
                    & 3 & 0 & 0 & 0 & 0 & 1 & 0 & \multicolumn{6}{l|}{}                         &                           \\ \cline{1-9}\cline{15-15}
\multirow{3}{*}{7}  & 1 & 0 & 1 & 0 & 1 & 0 & 1 & 1 & \multicolumn{5}{l|}{}                     & \multirow{3}{*}{$380.4944$} \\ \cline{2-9}
                    & 2 & 1 & 0 & 0 & 0 & 1 & 0 & 0 & \multicolumn{5}{l|}{}                     &                           \\ \cline{2-9}
                    & 3 & 0 & 0 & 1 & 0 & 0 & 0 & 0 & \multicolumn{5}{l|}{}                     &                           \\ \cline{1-10}\cline{15-15}
\multirow{3}{*}{8}  & 1 & 0 & 1 & 0 & 1 & 0 & 1 & 1 & 1 & \multicolumn{4}{l|}{}                 & \multirow{3}{*}{$385.4990$} \\ \cline{2-10}
                    & 2 & 1 & 0 & 0 & 0 & 1 & 0 & 0 & 0 & \multicolumn{4}{l|}{}                 &                           \\ \cline{2-10}
                    & 3 & 0 & 0 & 1 & 0 & 0 & 0 & 0 & 0 & \multicolumn{4}{l|}{}                 &                           \\ \cline{1-11}\cline{15-15}
\multirow{3}{*}{9}  & 1 & 0 & 1 & 0 & 1 & 0 & 1 & 0 & 1 & 1 & \multicolumn{3}{l|}{}             & \multirow{3}{*}{$390.7592$} \\ \cline{2-11}
                    & 2 & 1 & 0 & 1 & 0 & 0 & 0 & 1 & 0 & 0 & \multicolumn{3}{l|}{}             &                           \\ \cline{2-11}
                    & 3 & 0 & 0 & 0 & 0 & 1 & 0 & 0 & 0 & 0 & \multicolumn{3}{l|}{}             &                           \\ \cline{1-12}\cline{15-15}
\multirow{3}{*}{10} & 1 & 0 & 1 & 0 & 1 & 0 & 1 & 0 & 1 & 0 & 1 & \multicolumn{2}{l|}{}         & \multirow{3}{*}{$385.6078$} \\ \cline{2-12}
                    & 2 & 1 & 0 & 1 & 0 & 0 & 0 & 1 & 0 & 0 & 0 & \multicolumn{2}{l|}{}         &                           \\ \cline{2-12}
                    & 3 & 0 & 0 & 0 & 0 & 1 & 0 & 0 & 0 & 1 & 0 & \multicolumn{2}{l|}{}         &                           \\ \cline{1-13}\cline{15-15}
\multirow{3}{*}{11} & 1 & 0 & 1 & 1 & 0 & 1 & 0 & 1 & 0 & 1 & 0 & 1 & \multicolumn{1}{l|}{}     & \multirow{3}{*}{$382.4858$} \\ \cline{2-13}
                    & 2 & 1 & 0 & 0 & 1 & 0 & 0 & 0 & 1 & 0 & 0 & 0 & \multicolumn{1}{l|}{}     &                           \\ \cline{2-13}
                    & 3 & 0 & 0 & 0 & 0 & 0 & 1 & 0 & 0 & 0 & 1 & 0 & \multicolumn{1}{l|}{}     &                           \\ \cline{1-14}\cline{15-15}
\multirow{3}{*}{12} & 1 & 0 & 1 & 0 & 1 & 0 & 1 & 0 & 1 & 0 & 1 & 0 & 1                         & \multirow{3}{*}{$385.3658$} \\ \cline{2-14} 
                    & 2 & 1 & 0 & 1 & 0 & 0 & 0 & 1 & 0 & 1 & 0 & 0 & 0                         &                           \\ \cline{2-14} 
                    & 3 & 0 & 0 & 0 & 0 & 1 & 0 & 0 & 0 & 0 & 0 & 1 & 0                         &                           \\ \hline
\end{tabular}
\label{tab:3plants1Channel}
\end{table*}

\section{Numerical Results} \label{sec:numerical_example}
In the first example, we consider a collection of three unstable linear subsystems sharing a single communication channel. The entries of $A^{(i)}\in\mathbb{R}^{n_{i}\times n_{i}}$, $B^{(i)}\in\mathbb{R}^{n_{i}\times m_{i}}$, $C^{(i)}\in\mathbb{R}^{p_{i}\times n_{i}}$, $Q^{(i)}\in\mathbb{S}_{\succeq 0}^{n_{i}}$, $R^{(i)}\in\mathbb{S}_{\succ 0}^{m_{i}}$, $W^{(i)}\in\mathbb{S}_{\succeq 0}^{n_{i}}$ and $V^{(i)}\in\mathbb{S}_{\succ 0}^{p_{i}}$ are sampled independently from $\mathrm{Uni}(0,1)$. The dimensions are set as $n_{i} = 2$, $m_{i} = 1$, and $p_{i} = 1$ for all $i\in\{1,2,3\}$. Via exhaustive search, we first determine the optimal period and the associated communication sequence as well as the corresponding control loss as listed in Table~\ref{tab:3plants1Channel}. As highlighted in Table~\ref{tab:3plants1Channel}, $T_{0}=7$ gives the lowest control loss. For sufficiently large computational budgets, the MCTS with $c_{uct}=1.2$ finds the communication sequences that provide the same control loss obtained by the exhaustive search. In this example, we set the maximum number of iterations, performed by the MCTS, as $40,000$ when $T_{0}=12$. Notice that the exhaustive search requires performing $3^{12}$ function evaluations.

In the second example, we consider a group of five unstable subsystems that communicate over a network with two channels. Similar to the previous example, the entries of the matrices are randomly sampled from the same distribution, and their dimensions are set as $n_{i} = 4$, $m_{i} = 3$ and $p_{i} = 2$ for all $i\in\{1,2,3,4,5\}$. After performing $150,000$ iterations, for $T_{0}=10$, the MCTS with $c_{uct} = 1.4$ finds the communication sequence, shown in Table~\ref{tab:example_2}. Its control loss is computed as $J_{\textrm{ave}} = 4108.4376$.

\begin{table*}[ht!]
\caption{Five feedback loops sharing two communication channels}
\centering
\begin{tabular}{|c|c| >{\centering}p{0.4cm} | >{\centering}p{0.4cm} | >{\centering}p{0.4cm} | >{\centering}p{0.4cm} | >{\centering}p{0.4cm} | >{\centering}p{0.4cm} | >{\centering}p{0.4cm} | >{\centering}p{0.4cm} | >{\centering}p{0.4cm} | >{\centering}p{0.4cm} |c|}
\hline
\multirow{3}{*}{Period $T_{0}$} & \multirow{3}{*}{Plant index $i$} & \multicolumn{10}{l|}{Periodic sequence $\big\{\sigma_{m + kT_{0}}^{(i)}\big\}_{k\in\mathbb{N}_{0}}, ~ 0\leq m <T_{0}$} & \multirow{3}{*}{Total loss $J_{\textrm{ave}}$} \\ \cline{3-12} 
& & \multicolumn{10}{c|}{$m$} & \\ \cline{3-12}
& & 0 & 1 & 2 & 3 & 4 & 5 & 6 & 7 & 8 & 9 & \\ \hline
\multirow{5}{*}{10}  & 1 & 0 & 0 & 1 & 0 & 0 & 0 & 1 & 0 & 0 & 1 & \multirow{5}{*}{$4108.4376$}     \\ \cline{2-12} 
                     & 2 & 0 & 1 & 0 & 1 & 0 & 1 & 0 & 0 & 1 & 0 &                                  \\ \cline{2-12} 
                     & 3 & 1 & 0 & 0 & 1 & 0 & 1 & 0 & 1 & 0 & 1 &                                  \\ \cline{2-12} 
                     & 4 & 1 & 0 & 1 & 0 & 1 & 0 & 1 & 0 & 1 & 0 &                                  \\ \cline{2-12} 
                     & 5 & 0 & 1 & 0 & 0 & 1 & 0 & 0 & 1 & 0 & 0 &                                  \\ \hline
\end{tabular}
\label{tab:example_2}
\end{table*}

The reader can download the code used to create the tables presented in Section~\ref{sec:numerical_example} from \url{https://github.com/demirelbu/periodic-schedules} and run them to reproduce the results.

\section{Conclusions} \label{sec:conclusion}
This paper considers a networked control system that consists of a multitude of independent feedback loops closed over a shared network with multiple channels. We employ a centralized scheduler that generates a set of periodic communication sequences for allocating all available channels. Under the periodic, exogenous scheduling decisions made by the centralized scheduler, we design an optimal output feedback controller and derive analytical expressions for quantifying the quadratic control loss. Finally, we find the length of the periodic schedules, which attains the lowest overall control loss using both the exhaustive search and the Monte Carlo tree search.

\section{Appendix} \label{sec:appendix}
\noindent\textbf{Proof of Lemma~\ref{lem:eventually_periodic_sequence}.}
Assume that $\{\sigma_{t}^{(i)}\}_{t\in\mathbb{N}_{0}}$ is $T_{0}$-periodic (i.e., $\sigma_{t+T_{0}}^{(i)}=\sigma_{t}^{(i)}, ~\forall t\in\mathbb{N}_{0}$) and there exists at least one $m\in\{0, 1, \cdots, T_{0}-1\}$ such that $\sigma_{m+kT_{0}}^{(i)}=1, ~\forall k\in\mathbb{N}_{0}$. As shown in~\eqref{eqn:timer}, $\{\tau_{t}^{(i)}\}_{t\in\mathbb{N}_{0}}$ is generated by $\{\sigma_{t}^{(i)}\}_{t\in\mathbb{N}_{0}}$, and the assumption mentioned above leads to $\max_{t\in\mathbb{N}_{0}}\tau_{t}^{(i)}<T_{0}$. Notice that $\sigma_{t}^{(i)}=1 \Rightarrow \tau_{t}^{(i)}=0,~\forall t\in\mathbb{N}_{0}$ by checking~\eqref{eqn:timer}. If $\tau_{k T_{0}}^{(i)}=\tau_{0}^{(i)},~\forall k\in\mathbb{N}_{0}$, then $\tau_{t+T_{0}}^{(i)}=\tau_{t}^{(i)},~\forall t\in\mathbb{N}_{0}$ (i.e., $T_{0}$-periodicity). This statement can be verified by the inspection of~\eqref{eqn:timer}. To verify the $T_{0}$-periodicity of $\{\tau_{t}^{(i)}\}_{t\in\mathbb{N}_{0}}$, we have to investigate the following three cases:
\begin{itemize}
    \item[\textbf{(a)}] Suppose that $\sigma_{k T_{0}}^{(i)}=1,~\forall k\in\mathbb{N}_{0}$. Then, $\sigma_{k T_{0}}^{(i)}=\sigma_{0}^{(i)}=1 \implies \tau_{k T_{0}}^{(i)}=\tau_{0}^{(i)}=0,~\forall k\in\mathbb{N}_{0}$. 
    \item[\textbf{(b)}] Suppose that $\sigma_{k T_{0}}^{(i)}=0$ and $\sigma_{(k+1)T_{0}-1}^{(i)}=1,~\forall k\in\mathbb{N}_{0}$. Then, $\sigma_{k T_{0}}^{(i)}=\sigma_{0}^{(i)}=0 \implies \tau_{k T_{0}}^{(i)}=\tau_{0}^{(i)}=1, ~\forall k\in\mathbb{N}_{0}$ since $\tau_{t}^{(i)} = 0,~\forall t < 0$ and $\tau_{(k+1)T_{0}-1}^{(i)}=0,~\forall k\in\mathbb{N}_{0}$.
    \item[\textbf{(c)}] Suppose that $\sigma_{k T_{0}}^{(i)}=0$ and $\sigma_{(k+1)T_{0}-1}^{(i)}=0,~\forall k\in\mathbb{N}_{0}$. Thus, there is at least one $m\in\{1, \cdots, T_{0}-2\}$ such that $\sigma_{m + k T_{0}}^{(i)} = 1,~\forall k\in\mathbb{N}_{0}$. Define $\bar{m}\triangleq\max\big\{ m\in\{1, \cdots, T_{0}-2\} \mid \sigma_{m+k T_{0}}^{(i)} = 1, ~\forall k\in\mathbb{N}_{0}\big\}$. Then, $\sigma_{k T_{0}}^{(i)}=\sigma_{0}^{(i)}=0 \centernot\implies \tau_{k T_{0}}^{(i)}=\tau_{0}^{(i)},~\forall k\in\mathbb{N}_{0}$ since $\tau_{0}^{(i)}=1$ while $\tau_{k T_{0}}^{(i)}=T_{0}-\bar{m}, ~\forall k\in\mathbb{N}$.
\end{itemize}
For the cases (a) and (b), $\{\tau_{t}^{(i)}\}_{t\in\mathbb{N}_{0}}$ is $T_{0}$-periodic, whereas, for the case (c), $\{\tau_{t}^{(i)}\}_{t\in\mathbb{N}_{0}}$ is not $T_{0}$-periodic. For the case (c), when $T_{0}\leq t\in\mathbb{N}_{0}$, $\{\tau_{t}^{(i)}\}_{t\in\mathbb{N}_{0}}$ becomes $T_{0}$-periodic because $\tau_{k T_{0}}^{(i)} = \tau_{T_{0}}^{(i)} = T_{0}-\bar{m}, ~\forall k\in\mathbb{N}$. Hence, it is said to be eventually $T_{0}$-periodic. This concludes the proof. \hfill$\square$

\noindent\textbf{Proof of Lemma~\ref{lem:asymptotically_periodic_sequence}}
Assume that $\{\sigma_{t}^{(i)}\}_{t\in\mathbb{N}_{0}}$ is $T_{0}$-periodic (i.e., $\sigma_{t+T_{0}}^{(i)}=\sigma_{t}^{(i)}, ~\forall t\in\mathbb{N}_{0}$) and there is at least one $m\in\{0, 1, \cdots, T_{0}-1\}$ such that $\sigma_{m+kT_{0}}^{(i)}=\sigma_{m}^{(i)}=1,~\forall k\in\mathbb{N}_{0}$. According to Lemma~\ref{lem:eventually_periodic_sequence}, $\{\tau_{t}^{(i)}\}_{t\in\mathbb{N}_{0}}$ is $T_{0}$-periodic (or eventually $T_{0}$-periodic when $\sigma_{k T_{0}}^{(i)}=\sigma_{(k+1)T_{0}-1}^{(i)}=0,~\forall k\in\mathbb{N}_{0}$). As described in Remark~\ref{rem:periodic_sequence}, $\{\tau_{t}^{(i)}\}_{t\in\mathbb{N}_{0}}$ is $T_{0}$-periodic after a pre-period of length $T_{0}$.

As stated in~\cite[Theorem~4.1]{CGS:84}, for any given initial condition $P_{0\mid 0}^{s(i)}\in\mathbb{S}_{\succeq 0}^{n_{i}}$, there exists a constant matrix $P_{\infty}^{s(i)}\in\mathbb{S}_{\succeq 0}^{n_{i}}$, which is the stabilizing solution of the discrete-time Algebraic Riccati Equation, such that $\lim_{t\rightarrow\infty}P_{t\mid t}^{s(i)} = P_{\infty}^{s(i)}$ since the pairs $(A_{}^{(i)},W_{}^{(i)\nicefrac{1}{2}})$ are controllable and the pairs $(A_{}^{(i)},C_{}^{(i)})$ are observable for any given $i\in [N]$. Since $\lim_{t\rightarrow\infty}P_{t\mid t}^{s(i)} = P_{\infty}^{s(i)}$ holds, $\lim_{t\rightarrow\infty}K_{t}^{(i)} = K_{\infty}^{(i)}$ holds. Since both $\lim_{t\rightarrow\infty}P_{t\mid t}^{s(i)} = P_{\infty}^{s(i)}$ and  $\lim_{t\rightarrow\infty}K_{t}^{(i)} = K_{\infty}^{(i)}$ hold, $\lim_{t\rightarrow\infty}\Pi_{t}^{(i)} = \Pi_{\infty}^{(i)}$ holds.

Let $\{\Sigma_{t}^{(i)}\}_{t\in\mathbb{N}_{0}}$ be a sequence of positive semi-definite matrices, i.e.,
\begin{align*}
	\Sigma_{t}^{(i)} =  
	\begin{cases}
		\mathbf{0}_{n_{i}} & \text{if} ~ \tau_{t}^{(i)} = 0 \;, \\
		\sum\limits_{j=0}^{\tau_{t}^{(i)}-1}(A_{}^{(i)})^{j}\Pi_{t-1-j}^{(i)}(A_{}^{(i)})^{j\top} & \text{otherwise} \;.
	\end{cases} 
\end{align*}

Let $\{\bar{\Sigma}_{t}^{(i)}\}_{t\in\mathbb{N}_{0}}$ be a sequence of positive semi-definite matrices, i.e.,
\begin{align*}
	\bar{\Sigma}_{t}^{(i)} =  
	\begin{cases}
		\mathbf{0}_{n_{i}} & \text{if} ~ \tau_{t}^{(i)} = 0 \;, \\
		\sum\limits_{j=0}^{\tau_{t}^{(i)}-1}(A_{}^{(i)})^{j}\Pi_{\infty}^{(i)}(A_{}^{(i)})^{j\top} & \text{otherwise} \;,
	\end{cases} 
\end{align*}
which is $T_{0}$-periodic (or eventually $T_{0}$-periodic when $\sigma_{k T_{0}}^{(i)}=\sigma_{(k+1)T_{0}-1}^{(i)}=0,~\forall k\in\mathbb{N}_{0}$) since $\Pi_{\infty}^{(i)}\in\mathbb{S}_{\succeq 0}^{n_{i}}$ is constant and $\{\tau_{t}^{(i)}\}_{t\in\mathbb{N}_{0}}$ is $T_{0}$-periodic (or eventually $T_{0}$-periodic when $\sigma_{k T_{0}}^{(i)}=\sigma_{(k+1)T_{0}-1}^{(i)}=0,~\forall k\in\mathbb{N}_{0}$).

Define $\{\Sigma_{t}^{(i)} - \bar{\Sigma}_{t}^{(i)}\}_{t\in\mathbb{N}_{0}}$ whose terms are given by
\begin{multline}
	\Sigma_{t}^{(i)} - \bar{\Sigma}_{t}^{(i)} = \\
	\begin{cases}
		\mathbf{0}_{n_{i}} & \text{if} ~ \tau_{t}^{(i)} = 0 \;, \\
		\sum\limits_{j=0}^{\tau_{t}^{(i)}-1}(A_{}^{(i)})^{j}\Delta\Pi_{t-1-j}^{(i)}(A_{}^{(i)})^{j\top} & \text{otherwise} \;,
	\end{cases}
	\label{eqn:difference_sequence}
\end{multline}
where $\Delta\Pi_{t-1-j}^{(i)}\triangleq \Pi_{t-1-j}^{(i)}-\Pi_{\infty}^{(i)}$.

By inspection of~\eqref{eqn:difference_sequence}, it is trivial that if $\tau_{t}^{(i)} = 0, ~\forall t\in\mathbb{N}_{0}$, then $\lim_{t\rightarrow\infty}\Sigma_{t}^{(i)} - \bar{\Sigma}_{t}^{(i)} = \mathbf{0}_{n_{i}}$. If $\tau_{t}^{(i)} \neq 0, ~\forall t\in\mathbb{N}_{0}$, then $\lim_{t\rightarrow\infty}\Sigma_{t}^{(i)} - \bar{\Sigma}_{t}^{(i)} = \mathbf{0}_{n_{i}}$ still holds since $\lim_{t\rightarrow\infty}\Pi_{t-1-j}^{(i)}-\Pi_{\infty}^{(i)} = \mathbf{0}_{n_{i}}$. By Definition~\ref{def:asymptotically_periodic}, $\{\Sigma_{t}^{(i)}\}_{t\in\mathbb{N}_{0}}$ is asymptotically $T_{0}$-periodic because $\{\bar{\Sigma}_{t}^{(i)}\}_{t\in\mathbb{N}_{0}}$ is $T_{0}$-periodic (or eventually $T_{0}$-periodic when $\sigma_{k T_{0}}^{(i)}=\sigma_{(k+1)T_{0}-1}^{(i)}=0,~\forall k\in\mathbb{N}_{0}$) and $\lim_{t\rightarrow\infty}\Sigma_{t}^{(i)} - \bar{\Sigma}_{t}^{(i)} = \mathbf{0}_{n_{i}}$. 
This concludes the proof. \hfill$\square$

\begin{lemma}[Ces\`{a}ro means]\label{lem:Cesaro}
    Let $a_{n}\rightarrow a$ and let $b_{n}=n^{-1}\sum_{k=0}^{n}a_{k}$, then $\lim_{n\rightarrow\infty}b_{n} = a$.
\end{lemma}

\noindent\textbf{Proof of Theorem~\ref{thm:control_performance}.}
Here, we only focus on the proof of \emph{(iv)} since the proof of \emph{(i)}, \emph{(ii)}, and \emph{(iii)} can be found in~\cite{Ast:06}. As described in~\cite{Ast:06}, the expected minimum infinite-horizon control loss can be obtained as
\begin{multline}
		J_{\textrm{ave}}^{(i)} = \lim_{T\rightarrow\infty}^{}\frac{1}{T}J_{T}^{(i)} = \mathrm{Tr}\big( S_{\infty}^{(i)}W_{}^{(i)} \big) + \mathrm{Tr}\big( F_{\infty}^{s(i)}\Gamma_{\infty}^{(i)} \big) \\
		+ \lim_{T\rightarrow\infty}^{}\frac{1}{T}\sum_{t=0}^{T-1}\mathbf{E}\left[ \e_{t\mid t}^{(i)\top}\Gamma_{t}^{(i)}\e_{t\mid t}^{(i)} \right] \;.
		\label{eqn:inf_horizon_control_loss}
\end{multline}
The last term of~\eqref{eqn:inf_horizon_control_loss} can be re-written as
\begin{align}
    \lim_{T\rightarrow\infty}^{}\frac{1}{T}\sum_{t=0}^{T-1}&\mathbf{E}\left[ \e_{t\mid t}^{(i)\top}\Gamma_{t}^{(i)}\e_{t\mid t}^{(i)} \right] 
    = \lim_{T\rightarrow\infty}^{}\frac{1}{T}\sum_{t=0}^{T-1}\mathrm{Tr}\big( \Gamma_{\infty}^{(i)}\bar{\Sigma}_{t}^{(i)} \big) \nonumber\\ 
    &+ \lim_{T\rightarrow\infty}^{}\frac{1}{T}\sum_{t=0}^{T-1}\mathrm{Tr}\left( \Gamma_{\infty}^{(i)}(\Sigma_{t}^{(i)}-\bar{\Sigma}_{t}^{(i)}) \right) \nonumber\\ 
    &+ \lim_{T\rightarrow\infty}^{}\frac{1}{T}\sum_{t=0}^{T-1}\mathrm{Tr}\left( (\Gamma_{t}^{(i)}-\Gamma_{\infty}^{(i)})\Sigma_{t}^{(i)} \right) \;.
    \label{eqn:inf_horizon_control_loss_v2}
\end{align}
To complete the proof, we investigate three cases below.

\textbf{(a)} Let us first assume that $\{\sigma_{t}^{(i)}\}_{t\in\mathbb{N}_{0}}$ is $T_{0}$-periodic and there is at least one $m\in\{0, 1, \cdots, T_{0}-1\}$ such that $\sigma_{m+kT_{0}}^{(i)}=1,~\forall k\in\mathbb{N}_{0}$. Therefore, by Lemma~\ref{lem:asymptotically_periodic_sequence}, $\{\Sigma_{t}^{(i)}\}_{t\in\mathbb{N}_{0}}$ is an asymptotically $T_{0}$-periodic sequence that converges element-wise to $\{\bar{\Sigma}_{t}^{(i)}\}_{t\in\mathbb{N}_{0}}$. It is worth noting that if $\sigma_{k T_{0}}^{(i)}=\sigma_{(k+1)T_{0}-1}^{(i)}=0, ~\forall k\in\mathbb{N}_{0}$, then $\{\bar{\Sigma}_{t}^{(i)}\}_{t\in\mathbb{N}_{0}}$ is eventually $T_{0}$-periodic. Otherwise, $\{\bar{\Sigma}_{t}^{(i)}\}_{t\in\mathbb{N}_{0}}$ is $T_{0}$-periodic. As stated in~\cite[Theorem~4.1]{CGS:84}, for any given initial condition $S_{0}^{(i)}\in\mathbb{S}_{\succeq 0}^{n_{i}}$, there exists a constant matrix $S_{\infty}^{(i)}\in\mathbb{S}_{\succeq 0}^{n_{i}}$, which the stabilizing solution of the discrete-time Algebraic Riccati Equation, because the pairs $(A_{}^{(i)},B_{}^{(i)})$ are controllable and the pairs $(A_{}^{(i)},Q_{}^{(i)\nicefrac{1}{2}})$ are observable for all $i\in [N]$. Therefore, $\Gamma_{t}^{(i)}$ converges element-wise to $\Gamma_{\infty}^{(i)}$. Since $\lim_{t\rightarrow\infty}\Sigma_{t}^{(i)}-\bar{\Sigma}_{t}^{(i)}=\mathbf{0}_{n_{i}}$ and  $\lim_{t\rightarrow\infty}\Gamma_{t}^{(i)}-\Gamma_{\infty}^{(i)}=\mathbf{0}_{n_{i}}$, according to Lemma~\ref{lem:Cesaro}, the second and third terms of~\eqref{eqn:inf_horizon_control_loss_v2} become zero.

Let $a_{t}^{(i)}=\mathrm{Tr}\left( \Gamma_{\infty}^{(i)}\bar{\Sigma}_{t}^{(i)} \right)$. Notice that $a_{t}^{(i)}\geq 0$ due to $\Gamma_{\infty}^{(i)}, \bar{\Sigma}_{t}^{(i)}\in\mathbb{S}_{\succeq 0}^{n_{i}}$. Therefore, $\sum_{t=0}^{T-1}a_{t}^{(i)}$ is non-decreasing. Since $\{\bar{\Sigma}_{t}^{(i)}\}_{t\in\mathbb{N}_{0}}$ is $T_{0}$-periodic (or eventually $T_{0}$-periodic when $\sigma_{k T_{0}}^{(i)}=\sigma_{(k+1)T_{0}-1}^{(i)}=0,~\forall k\in\mathbb{N}_{0}$), $\{a_{t}^{(i)}\}_{t\in\mathbb{N}_{0}}$ is similarly $T_{0}$-periodic (or eventually $T_{0}$-periodic when $\sigma_{k T_{0}}^{(i)}=\sigma_{(k+1)T_{0}-1}^{(i)}=0,~\forall k\in\mathbb{N}_{0}$).

Assume $T\geq T_{0}$, then we have $T = \tilde{t}T_{0}^{}+r_{T}^{}$ with $0\leq r_{T}^{} \leq T_{0}^{}$ for some $\tilde{t}\in\mathbb{N}$. By the definition of $T$, $\tilde{t}T_{0} \leq T \leq (\tilde{t}+1)T_{0}$ holds for every $\tilde{t}\in\mathbb{N}$. Hence, we have:
\begin{align}
    \sum_{t=0}^{\tilde{t}T_{0}^{}-1}a_{t}^{(i)} \leq \sum_{t=0}^{T-1}a_{t}^{(i)}
    \leq \sum_{t=0}^{(\tilde{t}+1)T_{0}^{}-1}a_{t}^{(i)} \;,
    \label{eqn:inequality_v1}
\end{align}
or equivalently, 
\begin{align}
    c + \sum_{t=T_{0}}^{\tilde{t}T_{0}^{}-1}a_{t}^{(i)} \leq \sum_{t=0}^{T-1}a_{t}^{(i)}
    \leq c + \sum_{t=T_{0}}^{(\tilde{t}+1)T_{0}^{}-1}a_{t}^{(i)} \;,
    \label{eqn:inequality_v2}
\end{align}
where $c=\sum_{t=0}^{T_{0}-1}a_{t}^{(i)}$. Since $\{a_{t}^{(i)}\}_{t\in\mathbb{N}_{0}}$ is $T_{0}$-periodic after a pre-period of $T_{0}$ even if $\sigma_{k T_{0}}^{(i)}=\sigma_{(k+1)T_{0}-1}^{(i)}=0, ~\forall k\in\mathbb{N}_{0}$, \eqref{eqn:inequality_v2} is equal to
\begin{align}
    c + (\tilde{t}-1)\sum_{t=T_{0}}^{2T_{0}^{}-1}a_{t}^{(i)} \leq \sum_{t=0}^{T-1}a_{t}^{(i)} \leq c + \tilde{t}\sum_{t=T_{0}}^{2T_{0}^{}-1}a_{t}^{(i)} \;.
    \label{eqn:inequality_v3}
\end{align}
Taking the reciprocal of both sides of $\tilde{t}T_{0} \leq T \leq (\tilde{t}+1)T_{0}$, we get:
\begin{align}
    \frac{1}{(\tilde{t}+1)T_{0}^{}} \leq \frac{1}{T} \leq \frac{1}{\tilde{t} T_{0}^{}} \;.
    \label{eqn:inequality_v4}
\end{align}
Since $a_{t}^{(i)}\geq 0$ and $\sum_{t=0}^{T}a_{t}^{(i)}$ is non-decreasing, we can combine \eqref{eqn:inequality_v3} and \eqref{eqn:inequality_v4} as
\begin{multline}
    \frac{c}{(\tilde{t}+1)T_{0}}+
    \frac{\tilde{t}-1}{\tilde{t}+1}\frac{1}{T_{0}}\sum_{t=T_{0}}^{2T_{0}-1}a_{t}^{(i)} \leq \frac{1}{T}\sum_{t=0}^{T-1}a_{t}^{(i)} \\ 
    \leq \frac{c}{\tilde{t}T_{0}}+\frac{1}{T_{0}}\sum_{t=T_{0}}^{2T_{0}-1}a_{t}^{(i)} \;.
\end{multline}
Since $\frac{\tilde{t}-1}{\tilde{t}+1}\rightarrow 1$, $\frac{1}{\tilde{t}+1}\rightarrow 0$ and $\frac{1}{\tilde{t}}\rightarrow 0$ as $T\rightarrow\infty$, we obtain:
\begin{align*}
    \lim_{T\rightarrow\infty}^{}\frac{1}{T}\sum_{t=0}^{T-1}\mathrm{Tr}\left( \Gamma_{\infty}^{(i)}\bar{\Sigma}_{t}^{(i)} \right) = \frac{1}{T_{0}^{}}\sum_{t=T_{0}}^{2T_{0}^{}-1}\mathrm{Tr}\left( \Gamma_{\infty}^{(i)}\bar{\Sigma}_{t}^{(i)} \right)\;.
\end{align*}

\textbf{(b)} Let us now assume that $\{\sigma_{t}^{(i)}\}_{t\in\mathbb{N}_{0}}$ is not periodic and $\lambda(A^{(i)}) < 1$. As $T\rightarrow\infty$, the non-zero term of \eqref{eqn:inf_horizon_control_loss_v2} becomes
\begin{multline*}
    \lim_{T\rightarrow\infty}^{}\frac{1}{T}\sum_{t=0}^{T-1}\mathbf{E}\left[ \e_{t\mid t}^{(i)\top}\Gamma_{t}^{(i)}\e_{t\mid t}^{(i)} \right] \\
    = \lim_{T\rightarrow\infty}^{}\frac{1}{T}\sum_{t=0}^{T-1}\mathrm{Tr}\bigg( \Gamma_{\infty}^{(i)}\underbrace{\sum_{j=0}^{t}(A_{}^{(i)})^{j}\Pi_{\infty}^{(i)}(A_{}^{(i)})^{j\top}}_{Z_{t}^{(i)}} \bigg)\;.
\end{multline*}
Since $\lambda(A^{(i)}) < 1$ and $\Pi_{\infty}^{(i)}\in\mathbb{S}_{\succeq 0}^{n_{i}}$, $Z_{t}^{(i)}$ converges to
\begin{align*}
    Z_{\infty}^{(i)} = \sum_{j=0}^{\infty}(A_{}^{(i)})^{j}\Pi_{\infty}^{(i)}(A_{}^{(i)})^{j\top}\;,
\end{align*}
which is the unique solution of $A^{(i)}Z_{\infty}^{(i)}A^{(i)\top} -  Z_{\infty}^{(i)} + \Pi_{\infty}^{(i)} = 0$. Since $Z_{t}^{(i)}\rightarrow Z_{\infty}^{(i)}$, by Lemma~\ref{lem:Cesaro}, we get:
\begin{align*}
    \lim_{T\rightarrow\infty}^{}\frac{1}{T}\sum_{t=0}^{T-1}\mathbf{E}\left[ \e_{t\mid t}^{(i)\top}\Gamma_{t}^{(i)}\e_{t\mid t}^{(i)} \right]
    = \mathrm{Tr}\left( \Gamma_{\infty}^{(i)}Z_{\infty}^{(i)} \right)\;.
\end{align*}

\textbf{(c)} Let us assume that $\{\sigma_{t}^{(i)}\}_{t\in\mathbb{N}_{0}}$ is not periodic and $\lambda(A^{(i)}) \geq 1$. Then, $Z_{t}^{(i)}$ diverges as $t\rightarrow\infty$. In return, $J_{\text{ave}}^{(i)}$ also diverges.
This concludes the proof. \hfill$\square$

\bibliographystyle{IEEEtran}
\bibliography{bibliography}

\end{document}